\documentclass{amsart}
\usepackage{amsmath,amsthm,amssymb,IMjournal}

\usepackage{graphicx}
\usepackage[dvips]{epsfig}
\usepackage{xcolor}

\begin{document}
\newtheorem{The}{Theorem}[section]

\numberwithin{equation}{section}

\title{Energy dissipation and hysteresis cycles in
pre-sliding transients of kinetic friction}

\author{Michael Ruderman}

\rec {Dec 2022. Author accepted manuscript, Sep 2023}

\abstract

The problem of transient hysteresis cycles induced by the
pre-sliding kinetic friction is relevant for analyzing the system
dynamics e.g. of micro- and nano-positioning instruments and
devices and their controlled operation. The associated energy
dissipation and consequent convergence of the state trajectories
occur due to the structural hysteresis damping of contact surface
asperities during reversals, and it is neither exponential (i.e.
viscous type) nor finite-time (i.e. Coulomb type). In this paper,
we discuss the energy dissipation and convergence during the
pre-sliding cycles and show how a piecewise smooth
force-displacement hysteresis map enters into the energy balance
of an unforced system of the second order. An existing friction
modeling approach with a low number of the free parameters, the
Dahl model, is then exemplified alongside the developed analysis.
\endabstract




\section{Introduction}\label{sec1}

\begin{figure}[!h]
\centering (a) \includegraphics[width=0.35\columnwidth]{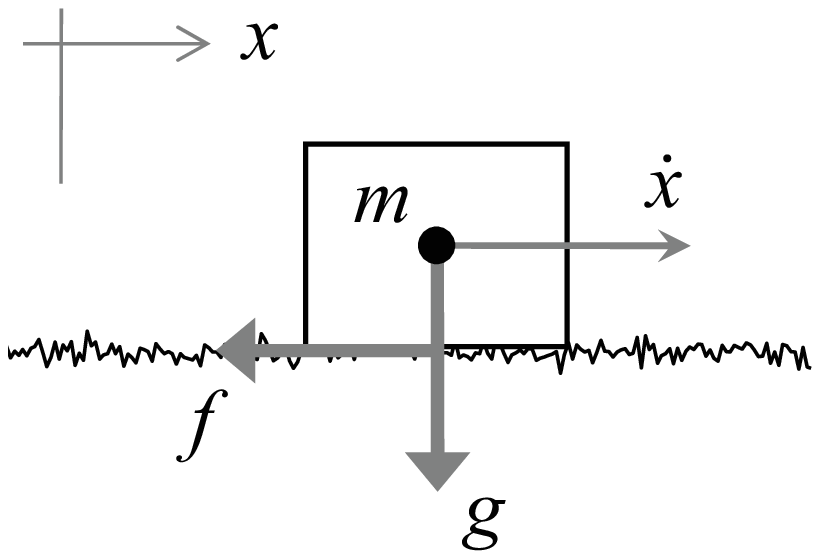}
\hspace{10mm} (b)
\includegraphics[width=0.4\columnwidth]{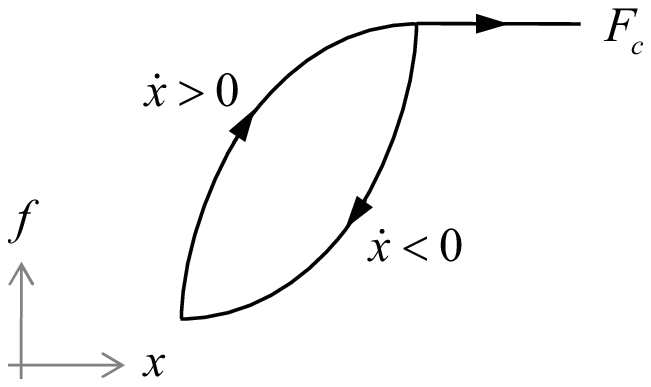} \caption{Schematic
representation of kinetic friction (a), pre-sliding friction
transitions in force-displacement plane (b).} \label{fig:1}
\end{figure}

The kinetic friction force in a small displacement range just
after the velocity reversals (that we call simply \emph{reversals}
in the following) is known to be dominated by the
pre-sliding/pre-rolling friction. During such pre-sliding
transitions, the classical Coulomb friction law can be extended by
a rate-independent hysteresis function, cf. e.g.
\cite{koizumi1984}. On the one hand, it allows to circumvent the
well-known and rather spurious discontinuity of the Coulomb
friction at zero velocity which, otherwise, complicates the
analysis of a system and may require solutions (e.g. in Filippov's
sense) of the differential inclusions, instead of differential
equations when describing the system dynamics. On the other hand,
the modeling of kinetic friction with pre-sliding hysteresis
reproduces better the measurable tribological phenomena known from
engineering practice, see e.g. in
\cite{koizumi1984,armstrong1994,lampaert2004}. The pre-sliding
frictional hysteresis manifests itself in the nonlinear
force-displacement characteristics with memory, cf. Figure
\ref{fig:1} (b), where each change in the velocity sign gives rise
to a new hysteresis branch. This can be compared with a so-called
''hysteretic spring'', cf. \cite{AlBender2004}, in which the
restoring force is not direction-symmetrical and acts as a source
of structural damping. The associated hysteresis losses in rolling
and sliding friction were recognized already in the early studies,
see e.g. \cite{greenwood1961}, and addressed both with various
theoretical approaches and experimentally. A remarkable
tribological study \cite{koizumi1984} has established several
rolling friction force-displacement relationships, especially with
regard to the area of hysteresis loops and rolling distance, and
the energy dissipated at contact surfaces. Recall that an
irregular rough surface which, at the same time, can be
characterized in terms of an average high, stiffness, density, and
distribution of asperities, appears as a structural source of
kinetic friction $f$, cf. Figure \ref{fig:1} (a).
The latter is opposed to direction of the relative displacement,
indicated by the velocity $\dot{x}$, and is usually proportional
to a normal load $g$ (caused by a point-mass $m$ as in the
picture). Despite the hysteresis dissipation properties are well
known and, so to say, typical for structural mechanics, see e.g.
\cite{lacarbonara2003} and the references therein, and were
largely covered also in the mathematical studies of hysteresis
systems, see e.g. \cite{Krejci96,BrokateSprekels96}, the
hysteresis in frictional systems was also strongly accented in the
system and control studies, see e.g.
\cite{armstrong1994,bliman1992mathematical}.

Several empirical and phenomenological dynamic models of kinetic
friction are known, most of which capture the hysteresis during
pre-sliding in often different ways. Due to the use of nonlinear
differential equations and often distributed parameters and
dynamic states with switching, a rigorous analysis of the damping
properties and state trajectories for such models has been and
remains a non-trivial task. For instance, the hysteresis damping
of LuGre and Maxwell-slip friction models were addressed in
\cite{ruderman2015}, while in \cite{AlBender2004} a generic
approach of analyzing hysteretic friction was presented with use
of the \emph{Masing rules}, see e.g. in
\cite{BrokateSprekels96,bertotti2005science} for details. In
almost all frictional studies, an importance of extending the
classical Coulomb friction law by the pre-sliding transitions was
(at least on the edge) recognized. This comes, however, at the
cost of significantly more complex dynamics of the overall system,
for which even a simple one-parameter Coulomb friction law with
discontinuity poses a significant challenge in finding the
analytical solution for trajectories, cf. e.g. \cite{ruderman22}.

The Dahl model \cite{dahl1968solid} was originally proposed for
describing the transient behavior of the solid friction when it is
piecewise elastic rather than viscous. It was probably a first
attempt to circumvent the discontinuity in the classical Coulomb
law of dry friction at the reversals. The Dahl model was
widespread (also in industrial applications) for analyzing and
simulating the solid friction occurring in the ball-bearing
systems. Having only two tunable parameters, one of which is the
Coulomb friction coefficient, the Dahl model in its simplified
form appears suitable for a detailed analysis of hysteresis cycles
and the associated energy dissipation.

This paper aims at investigating the dynamic behavior of a
standard lumped-mass motion system subject to the pre-sliding
frictional damping. Being motivated by the previous studies
\cite{bliman1992mathematical}, \cite{AlBender2004}, our goal is to
provide a clear and straightforward relationship between the
energy balance of reversals and state trajectories during the
hysteresis cycles. The underlying motivation is to have a generic
methodology that can be used for different modeling approaches
when describing the rate-independent friction-force-displacement
characteristics. Having said that, the following materials present
the main results organized in two sections. In section \ref{sec2},
we discuss the unforced pre-sliding hysteresis oscillator, while
establishing the energy balance and proving the convergence of
hysteresis dissipation cycles. In section \ref{sec3}, we apply the
developed analysis and exemplify the associated calculus by using
the Dahl friction model. The conclusions are given in section
\ref{sec4}.

\section{Unforced pre-sliding hysteresis oscillator}\label{sec2}

The unforced motion dynamics within a pre-sliding frictional
range, i.e. in some neighborhood to the motion reversals
characterized by $\mathrm{sign}\left[\dot{x}(t^+)\right] \neq
\mathrm{sign}\left[\dot{x}(t)\right]$ and $-1 < f(t)/F_c < 1$,
where $\dot{x}(t^+)$ denotes the (limit) value of $x$ just after
changing stepwise the sign i.e. when the time argument approaches
$t$ from the right, can be described in the generic form as
\begin{equation}
m \frac{d^2x}{dt^2} + f[x] = 0, \quad \frac{dx}{dt}(0) \neq 0.
\label{eq:2:1}
\end{equation}
Here the inertial motion of the point mass $m$ is counteracted by
the nonlinear restoring force $f[\cdot]$ which includes the
frictional damping. Note that the counteracting friction force $f$
is written as function of the relative displacement $x$, since it
is modeling an irreversible process which depends on $x(t)$ and
its previous values, in particular after each motion reversal,
i.e. when $\mathrm{sign}\left[\dot{x}(t)\right]$ changes. Further
we stress that an unforced system response owing to nonzero
initial conditions, cf. \eqref{eq:2:1}, is in focus of our
discussion. The forced pre-sliding hysteresis oscillations, for
which the right-hand side of \eqref{eq:2:1} should otherwise
contain a known exogenous value $u(t)$ in the sense of an applied
external force, could extend the analysis presented here. However,
it would go beyond the scope of this work.

In the following, for the sake of brevity, we will use the
notation $\dot{o}$ for describing time derivative of $o$ with
respect to time $t$; correspondingly the double dot above a
variable, i.e. $\ddot{o}$, denotes the second-order derivative
with respect to time, etc. Rewriting \eqref{eq:2:1} with $\ddot{x}
= \dot{x} d \dot{x} (dx)^{-1}$ and integrating subsequently both
sides of the resulted equation yields
\begin{equation}
m \int \dot{x} d \dot{x} = - \int f[x] d x. \label{eq:2:2}
\end{equation}
Note that \eqref{eq:2:2} is valid only in time intervals where
$x(t)$ is monotone, that is assumed to be the case between any two
consecutive reversal points. This leads to the energy balance of
an unforced pre-sliding hysteresis oscillator, as the total
kinetic and potential plus dissipated energies are written as
\begin{equation}
\frac{1}{2} m \, \dot{x}^2 + \int \limits_{x_j}^{x_{j+1}} f[x] d x
= 0, \label{eq:2:3}
\end{equation}
for two successive \emph{reversal points} $x_j \neq x_{j+1}$ at
times $t_j < t_{j+1}$. Let us denote the first kinetic energy term
in the above equation by $E_k$, and the second energy term
associated with the restoring force by $E_f$. Depending on the
restoring force $f$, that is generally nonlinear and
history-dependent, the energy $E_f = E_p + E_d$ includes both, the
potential energy $E_p(x)$ and the dissipation energy $E_d(C)$
along a path $C$. Since for each reversion instant $t_{i}$, where
$i \in \mathbb{Z}^{+}$, the $\dot{x}(t_i)=0$ is valid, one can
analyze the energy balance of the reversal cycles by using the
$E_f$ energies only, i.e. setting $E_k$ to zero. Note that several
constitutive friction models with hysteresis (like the Dahl model
assumed in the following) does not have the structure which would
allow considering a dissipation rate $D \geq 0$ and, thus,
following the principles of thermodynamics. Therefore, a
meaningful energy balance can be evaluated only at the reversal
points. Following to that, one can show that the total energy
dissipated during one closed cycle $C = (x_i, x_{i+1}, x_{i+2})$,
where the first and last reversion points coincide with each other
i.e. $x_i = x_{i+2}$, is given by
\begin{equation}
E_d(C) = \oint \limits_{C} f[x] d x   = E_f(x_i,x_{i+1}) -
E_p(x_i) - E_f(x_{i+1},x_{i+2}) + E_p(x_{i+2}). \label{eq:2:4}
\end{equation}
Given $E_p(x_{i+2})= E_p(x_i)$ for a closed path $C$ of one
reversion cycle, the total net dissipated energy can be calculated
as
\begin{equation}
E_d = \int \limits_{x_i}^{x_{i+1}} f[x] d x + \int
\limits_{x_{i+1}}^{x_i} f[x] d x. \label{eq:2:5}
\end{equation}

\begin{figure}[!h]
\centering
\includegraphics[width=0.4\columnwidth]{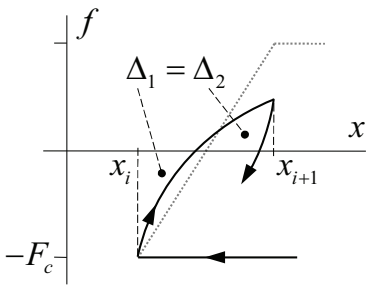}
\caption{Energy balanced reversion points in $(x,f)$ plane.}
\label{fig:2}
\end{figure}

First, let us briefly sketch the situation where the restoring
force would map a piecewise linear spring, i.e. $f \equiv
\bar{f}(x) = k \, x(t)$ on the interval $-1 < f(t)/F_c < 1$, with
some positive stiffness coefficient $0 < k < \infty$. Obviously,
the energy due to the restoring force will have only the potential
term, according to the Hooke's law, and is then expressed by $E_p
= 0.5 \, k x^2$. In this case, substituting $\bar{f}(x)$ into
\eqref{eq:2:5} and evaluating the definite integral yield
\begin{equation}
\bar{E}_d = \int \limits_{x_i}^{x_{i+1}} k x \, d x + \int
\limits_{x_{i+1}}^{x_i} k x \, d x = \frac{1}{2} kx^2 \Bigr
|_{x_i}^{x_{i+1}} - \frac{1}{2} kx^2 \Bigr |_{x_i}^{x_{i+1}} = 0.
\label{eq:2:6}
\end{equation}
The obtained result is intuitively logical since the linear
restoring force is purely conservative and does not dissipate (or
generate) any energy on the closed displacement cycles. This
situation is illustrated in the $(x,f)$ plane in Figure
\ref{fig:2} by the grey dot line. Note that once $|f(t)/F_c| \geq
1$, the $f$-force saturates at $\pm F_c$ level and, afterwards,
behaves as a constant Coulomb friction term. Recall that such
piecewise affine behavior can be well described by the
Prandtl-Ishlinskii stop-type operator \cite{Krejci96}, as shown
and discussed in \cite{ruderman2017b} in the context of the
Coulomb friction.

Now, assuming the restoring force is a nonlinear map with memory,
i.e. $f \equiv \hat{f}[x] \vee \check{f}[x]$ on the interval $-1 <
f(t)/F_c < 1$, where $\hat{f}$ is an ascending branch for
$\mathrm{sign}[\dot{x}]>0$ and $\check{f}$ is a descending branch
for $\mathrm{sign}[\dot{x}]<0$, one can equally use \eqref{eq:2:5}
for calculating the net dissipated energy at one closed reversion
cycle. More specifically, we assume $f[x]$ to be a clockwise
rate-independent hysteresis, see e.g. in
\cite{BrokateSprekels96,bertotti2005science} for basics on the
hysteresis operators. Under this assumption, one can show the
dissipated energy
\begin{equation}
\tilde{E}_d = \int \limits_{x_i}^{x_{i+1}} \Bigl( \hat{f}[x] -
\check{f}[x] \Bigr) \, d x = \Delta \neq 0. \label{eq:2:7}
\end{equation}
This follows from the fact that for a clockwise hysteresis map, an
ascending branch (in the time interval $t \in [t_i, t_{i+1}]$)
lies always above the descending branch (in the time interval $t
\in [t_{i+1}, t_{i+2}]$), given a pair of the input reversion
points $x_{i} < x_{i+1}$. Such situation is shown in Figure
\ref{fig:2} by the black solid line. The corresponding to
\eqref{eq:2:7} nonzero path integral
$$
\oint \limits_{C} f[x] d x = \Delta
$$
postulates the energy losses at one closed reversion cycle, i.e.
$E_d(C) = \Delta$, which is equal to the area enclosed by a
force-displacement loop, cf. Figure \ref{fig:1} (b). We will
denote it by \emph{hysteretic energy dissipation} and recall that,
in general, the amount of energy dissipated during a cycle is
proportional to the area of the enclosed hysteresis loop
\cite{BrokateSprekels96}. Important to notice is that the
corresponding damping of an oscillatory system is
rate-independent, similar to structural hysteresis damping that is
common in material science and structural mechanics. It is also
worth emphasizing that $\Delta \rightarrow 0$ if
$\bigl(\hat{f}[x_a] - \check{f}[x_b]\bigr) \rightarrow 0$ for all
$x_a=x_b$ on the intervals $x_i < x_a < x_{i+1}$ and $x_{i+1} <
x_b < x_{i+2}$ of a closed cycle $C$. That means, it is the
clockwise branching and formation of hysteresis loops that make
the restoring term $f[x]$ dissipative in \eqref{eq:2:1}.
Otherwise, a continuously increasing nonlinear but memory-free
restoring force function $f(x)$ would lead to zero energy
dissipation on the closed reversal cycles.

Now, we are in the position to address energy dissipation and
decaying hysteresis cycles from a viewpoint of the balance
equation \eqref{eq:2:3} obtained for the system \eqref{eq:2:1}.
Note that in the following, for the sake of simplicity and without
loss of generality, we will consider the unity mass, meaning
$m=1$.

Changing the variables of integration for the restoring force
energy, cf. \eqref{eq:2:3}, one can rewrite it as
\begin{equation}
E_f = \int \limits_{x_i}^{x_{i+1}} f[x] d x = \int
\limits_{t_i}^{t_{i+1}} f[x] \,\dot{x} \, d t. \label{eq:2:8}
\end{equation}
We consider an arbitrary pair of reversion points $x_i$ and
$x_{i+1}$ as shown in Figure \ref{fig:2}. Here the initial
condition for the system \eqref{eq:2:1} is $\dot{x}(0) < 0$; this
is without loss of generality. One can recognize that on the
interval $\left[x_i, x_0\right)$, where $x_0$ is the point of
$f(x_0)=0$, the integrand $f[\cdot]\, \dot{x} < 0$, since
$\mathrm{sign}(f) \neq \mathrm{sign}(\dot{x})$. On the opposite,
the integrand $f[\cdot] \, \dot{x} > 0$ on the interval
$\left(x_0, x_{i+1} \right]$, owing to the relative velocity and
the restoring force have the same sign. That is after an initial
reversal and until zero-crossing of the restoring force, the
restoring power is first negative, which implies the energy of
restoring force is converted into the kinetic one, cf. with
\eqref{eq:2:3}. As a consequence, the system \eqref{eq:2:1}
accelerates and reaches the peak relative velocity
$\dot{x}_{m}(t_0)$ at time $t_0$ of the force zero-crossing. The
peak relative velocity $|\dot{x}_{m}| = \max | \dot{x}(t) |$ is
the maximal or minimal (depending on direction of the last
reversal) on the interval $t_i \leq t \leq t_{i+1}$. Subsequently,
after force zero-crossing, the restoring power becomes positive
during $t \in \left(t_0, t_{i+1} \right]$, meaning the kinetic
energy $E_k$ is continuously converted into $E_f$ until reaching
the next reversal with $\dot{x}(t_{i+1}) = 0$. Since both
reversion points $x_{i}$ and $x_{i+1}$ are energetically balanced,
cf. \eqref{eq:2:3}, one can see that the corresponding integral
values have the same amount, i.e.
\begin{equation}
\int \limits_{t_i}^{t_{0}} f[x] \,\dot{x} \, d t + \int
\limits_{t_0}^{t_{i+1}} f[x] \,\dot{x} \, d t = 0 \label{eq:2:9}
\end{equation}
also meaning the equal areas $\Delta_1 = \Delta_2$, cf. Figure
\ref{fig:2}.

Now, let us demonstrate the convergence of such reversion cycles
for the system \eqref{eq:2:1}, provided $f[\cdot]$ is a clockwise
hysteresis map. Important to emphasize is that the first integral
in \eqref{eq:2:9} constitutes the potential energy which is
transferred into the kinetic one until $t=t_0$. In contrast, the
second integral in \eqref{eq:2:9} is the superposition of the
potential and dissipation energies, which are transferred from the
kinetic energy until the next reversal at $t=t_{i+1}$. The
potential energy part is the recoverable one when a new reversal
occurs. Since the potential energy has its maximum value at the
point of reversion, and the dissipation energy occurs on a way
between two consecutive reversion points, one can state
\begin{equation}
E_{p}(i) = E_{d}(i) + E_{p}(i+1). \label{eq:2:10}
\end{equation}
Here we are using the argument $i \in \mathbb{Z}^{+}$ for labeling
the reversion points. Realizing $E_{d}(i) \neq 0$ for all possible
hysteresis branches leads to a recursion expansion
$$
E_p(i) = E_d(i) + \underset{E_p(i+1)} {\underbrace{\Bigl( E_d(i+1)
+ E_p(i+2)  \Bigr)}} = E_d(i) + \Bigl( E_d(i+1) +  \bigl( \ldots
\underset{E_p(i+n)} {\underbrace{ (\ldots) }} \bigr) \Bigr),
$$
with $n \rightarrow \infty$. It means that after each reversal,
the amount of potential energy (cf. $\Delta$-areas in Figure
\ref{fig:2}) which can be then converted into kinetic one and thus
drives the oscillator trajectories becomes smaller and smaller. At
the same time, the total amount of the dissipated energy
increases. This leads to the series expansion
\begin{equation}
E_{p}(i) = \sum \limits_{n=i}^{\infty} E_{d}(n), \label{eq:2:11}
\end{equation}
which converges to the potential energy of the initial reversion
point, i.e. $E_{p}(i)$. Note that the initial potential energy is
bounded from above by
\begin{equation}
E_{p}(i) < \frac{1}{2 \, k_{\min}} \, F_c^2 , \label{eq:2:12}
\end{equation}
where $k_{\min}$ is the minimal possible equivalent stiffness of
the restoring force $f[\cdot]$, cf. with the case exemplified by
\eqref{eq:2:6}. This implies that the damped hysteresis
oscillations of the system \eqref{eq:2:1} are always amplitude
bounded in the $(x, \dot{x})$ coordinates. At the same time, the
damping rate an hence the state convergence, i.e. $\|(x, \dot{x})
\| \rightarrow 0$, depend strongly on the analytic form of the
hysteresis map $f[\cdot]$.

In the following, we apply the above developments to discuss in
details the pre-sliding oscillations based on one possible
modeling approach of capturing the Coulomb friction with
continuous pre-sliding transitions, namely the Dahl model
\cite{dahl1976}.

\section{Application with Dahl model}\label{sec3}

The original Dahl model in the differential form is given by, cf.
\cite{dahl1976},
\begin{equation}
\frac{d F(x)}{d x} = \sigma \Bigl | 1 - \frac{F}{F_c}
\mathrm{sgn}(\dot{x}) \Bigr |^\gamma \mathrm{sgn} \Bigl(  1 -
\frac{F}{F_c} \mathrm{sgn}(\dot{x})  \Bigr) , \label{eq:3:1}
\end{equation}
where $\mathrm{sgn}(\dot{x})$ captures the direction of relative
displacement, i.e. the sign of velocity. The rest stiffness $0 <
\sigma \sim (dF/dx)$ when $|F| \ll F_c$, cf.
\cite{bliman1992mathematical}, is a characteristic property of
asperities of the contact surface. The shaping factor $\gamma \geq
0$ permits modulation of the form of displacement-force curves
upon the reversion points. Worth mentioning is that for
$\gamma=0$, the expression \eqref{eq:3:1} becomes equivalent to
the Prandtl-Ishlinskii stop-type operator with piecewise linear
displacement-force transitions, cf. with \cite{ruderman2017b} and
example given around eq. \eqref{eq:2:6}. Also, if allowing for
$\sigma \rightarrow \infty$ with $\gamma=0$, the Dahl model
\eqref{eq:3:1} will reduce to the classical Coulomb one, that has
discontinuity at motion reversals, i.e. $F = F_c
\mathrm{sgn}(\dot{x})$. Since $|F(t)| \leq F_c$ for all times $0
\leq t < \infty$ it was proven in \cite{bliman1992mathematical}
that \eqref{eq:3:1} can be written in a simpler form
\begin{equation}
\frac{d F(x)}{d x} = \sigma \Bigl ( 1 - \frac{F}{F_c}
\mathrm{sgn}(\dot{x}) \Bigr )^\gamma. \label{eq:3:2}
\end{equation}
Also recall that for ductile type materials $\gamma \geq 1$, cf.
\cite{dahl1976}. In the sequel, we will assume $\gamma=1$ to keep
our analysis with the associated integral calculus simpler and
well tractable. It is also noteworthy that \eqref{eq:3:2} with
$\gamma=1$ is the most commonly used realization of the Dahl
model, widely used for systems and control studies.

With the assumptions made above, we can rewrite \eqref{eq:3:2} and
obtain
\begin{equation}
F_c \sigma^{-1} \,\frac{d F(x)}{d x} + F(x) \,
\mathrm{sgn}(\dot{x}) = F_c. \label{eq:3:3}
\end{equation}
The above expression reveals several interesting observations
about the Dahl model behavior. Namely, (i) the frictional force
$F(x)$ monotonically approaches the Coulomb friction level since
$|F| \rightarrow F_c$ after reversals, i.e. once the sign of
$\dot{x}$ changes. (ii) this transient behavior does not depend on
the time $t$ but on $x$ and, thus, represents a rate-independent
$F(x)$ map. (iii) the transition mapping $F(x)$ does not
explicitly depend on the coordinate of $x$ and is invariant to the
location or translation of the reversion points $x_i$, cf. section
\ref{sec2}. (iv) the convergence shape of $|F(x)| \rightarrow F_c$
depends on the ratio $\sigma/F_c$ between both frictional
parameters of the system.

Solving \eqref{eq:3:3}, with the initial condition $F_i(x_i)$
which corresponds to the last reversion point $x_i$, one obtains
the restoring force $f[\cdot] \equiv F\bigl(x, F_i(x_i)\bigr)$,
cf. section \ref{sec2}, that represents locally the Dahl model
written in the algebraic form
\begin{equation}
f[x] = \mathrm{sgn}(\dot{x}) \Bigl(F_c - \bigl(F_c -
\mathrm{sgn}(\dot{x}) F_i \bigr) \exp \bigl( -
\mathrm{sgn}(\dot{x}) \sigma F_{c}^{-1} (x-x_i) \bigr) \Bigr).
\label{eq:3:4}
\end{equation}
Since \eqref{eq:3:4} captures bidirectional progress of the
force-displacement curves, depending on $\mathrm{sgn}(\dot{x})$,
hereinafter we will consider the positive sign of velocity,
analogous to the situation shown in Figure \ref{fig:2} and without
loss of generality. Hence, obtaining
\begin{equation}
f_{+}[x] = F_c - \bigl(F_c - F_i \bigr) \exp \Bigl( -
\frac{\sigma}{F_c} (x-x_i) \Bigr) \label{eq:3:5}
\end{equation}
and first solving $f_{+}[x] = 0$ with respect to $x$, one can
determine
\begin{equation}
x_{0,i} = x_i - \frac{F_c}{\sigma}\, \ln \Bigl( \frac{F_c}{F_c -
F_i} \Bigr), \label{eq:3:6}
\end{equation}
which is the zero-crossing point of the restoring force, provided
$F_i < 0$, cf. with section \ref{sec2} and Figure \ref{fig:2}.

\begin{figure}[!h]
\centering
\includegraphics[width=0.8\columnwidth]{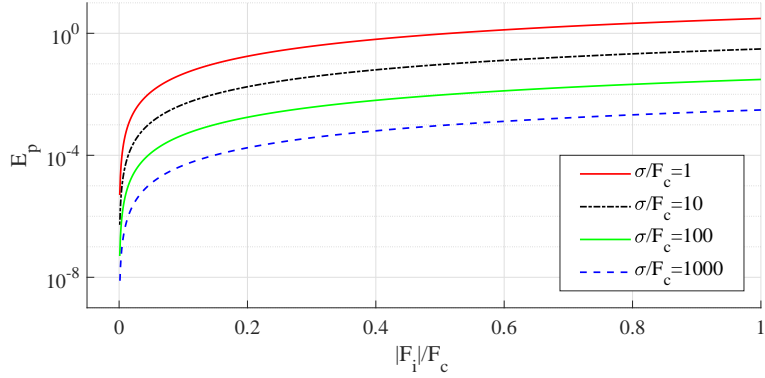}
\caption{Potential energy $E_p$ (on logarithmic scale) as a
function of the normalized restoring force $|F_i|/F_c$ at the
reversion point $x_i$.} \label{fig:3}
\end{figure}

Now, in order to analyze energy produced by the restoring force we
evaluate the antiderivative of \eqref{eq:3:5}, this with respect
to the initial conditions, and obtain
\begin{equation}
E_{f+}(x) = \int f_{+}[x] \, dx =  F_c \, x + \frac{F_c^2
}{\sigma} \exp \Bigl(-\frac{\sigma}{F_c} \, x \Bigr) - \frac{F_c^2
}{\sigma}. \label{eq:3:7}
\end{equation}
Worth noting is that the above result is consistent with analysis
of the energy stored by the hysteresis loops of the Dahl model
provided in \cite{bliman1992mathematical}. Recalling that
$E_{f+}(x)$ is defined only on the interval between two
consecutive reversion points, i.e. $x \in [x_i, x_{i+1}]$, and
using the fact of the balanced energy due to restoring force, cf.
\eqref{eq:2:9} and Figure \ref{fig:2}, one needs to consider
\eqref{eq:3:7} before and after zero-crossing point $x_0$. In
addition, for the correct sign of the integrands in the energy
balance \eqref{eq:2:9} and, thus, for taking into account the
initial conditions of integration \eqref{eq:3:7}, one has to
assign $x_i$ in relation to $x_{0,i}=0$. Out from \eqref{eq:3:6},
one obtains
\begin{equation}
x_i = \frac{F_c}{\sigma} \ln \Bigl(\frac{F_c}{F_c - F_i} \Bigr) <
0, \label{eq:3:8}
\end{equation}
which satisfies $-F_c \leq F_i < 0$. When evaluating the definite
integral of \eqref{eq:3:5}, with the set lower limit $x_0$ and
upper limit $x_i$, one obtains the potential energy as a function
of the restoring force at the reversion point:
\begin{equation}
E_p(F_i) = \frac{F_c^2 }{\sigma} \Biggl( \ln \Bigl( \frac{F_c}{F_c
- F_i} \Bigr) - \frac{F_i}{F_c} \Biggr). \label{eq:3:9}
\end{equation}
Note that \eqref{eq:3:9} is also equal to the maximal kinetic
energy over the interval $[t_i, t_{i+1}]$, i.e. during the motion
between the reversal points $x_i$ and $x_{i+1}$. Further we note
that the potential energy is bounded from above, cf. with
\eqref{eq:2:12}, by
$$
\max E_p = 0.3069 \, \frac{F_c^2 }{\sigma}
$$
since $F_i \in [-F_c,F_c]$. The potential energy \eqref{eq:3:9} is
visualized (on the logarithmic scale) in Figure \ref{fig:3} as a
function of the normalized restoring force $|F_i|/F_c$. The curves
are shown for the unitless ratio between the rest stiffness and
Coulomb friction bound of the Dahl model, i.e. $\sigma/F_c = \{1,
10, 100, 1000\}$.
One can recognize a considerable drop in the potential energy,
closer to zero is the reversal state in terms of the restoring
force. This serves also as an indication of how the potential
energy and the corresponding hysteresis oscillations will decrease
over the course of reversal cycles.

\begin{figure}[!h]
\centering
\includegraphics[width=0.9\columnwidth]{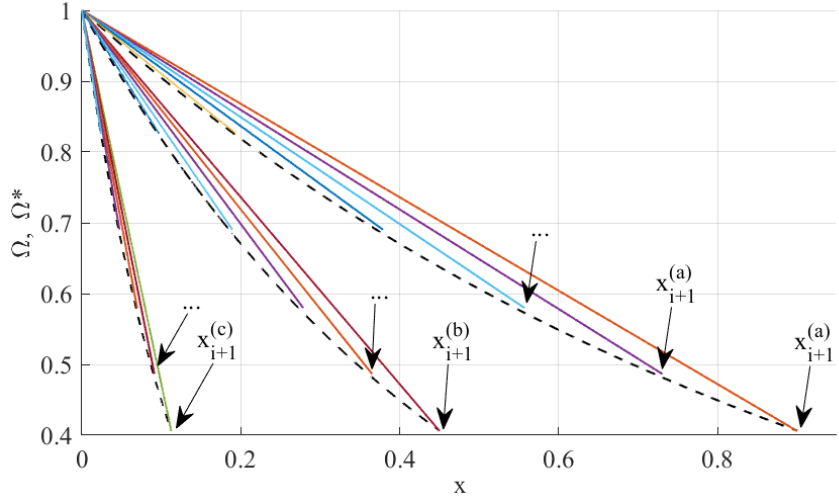}
\caption{Exponential term $\Omega$ (black dash lines) and its
linear approximation $\Omega^*$ (colored solid lines) for
different parameter ratios $\sigma/F_c^{(a,b,c)}= \{1, 2, 8\}$ and
initial states $F_i = -F_c \times \{0.2, 0.4, \ldots, 1\}$.}
\label{fig:4}
\end{figure}

\begin{figure}[!h]
\centering
\includegraphics[width=0.8\columnwidth]{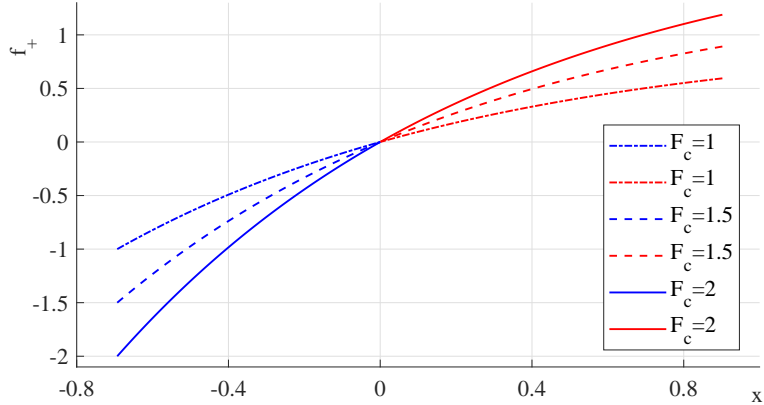}
\caption{Force-displacement curves between two reversion points
$i$ and $i+1$ for various Coulomb friction coefficients $F_c =
\{1, 1.5, 2\}$.} \label{fig:5}
\end{figure}

Once the potential energy \eqref{eq:3:9} is known, with respect to
the last reversal in $x_i$, we are interested to evaluate the
restoring force energy which is balanced by the kinetic energy
$E_k(t) = 0.5 m \dot{x}^2$. Recall that the kinetic energy is
first continuously increasing until reaching its maximum value,
i.e. with $\max |\dot{x}|$ at $t_0$, cf. section \ref{sec2}, and
then continuously decreasing until zero velocity, that gives rise
to the next reversal in $x_{i+1}$. This is reflected in the energy
balance of the restoring force, implying $E_f(x_{i+1}) = E_k(t_0)
= E_p(F_i)$. One can recognize that in order to determine
$x_{i+1}$ from \eqref{eq:3:7}, one needs to solve
\begin{equation}
\frac{F_c^2 }{\sigma} \, \Omega(x) + F_c x -\frac{F_c^2 }{\sigma}
= E_{f+}(x_i) \quad \hbox{ with } \quad \Omega(x) = \exp \Bigl( -
\frac{\sigma}{F_c} x \Bigr) \label{eq:3:10}
\end{equation}
with respect to $x$. The above equation has the unknown variable
$x$ both outside and inside the exponential function $\Omega$ and,
thus, cannot be solved explicitly. Recall that a non-explicit
solution of \eqref{eq:3:10} would require the multivalued Lambert
\emph{W}-function, see e.g. \cite{corless1996} for details. Due to
multivaluedness with complex numbers the Lambert \emph{W}-function
can hardly serve us to find the unique $x_{i+1} \in \mathbb{R}^+$.
Instead, we are deriving a suitable linear approximation
$\Omega^*$ which is sufficiently close to $\Omega$ on the interval
$x \in [0, x_{i+1}]$. This should also take into account the
parametric ratio $\sigma / F_c$ and the last reversal state, which
is mapped through $F_i$ and affects the energy $E_p$, cf.
\eqref{eq:3:9}. Requiring $\Omega(0) = \Omega^*(0)=1$ and
$\Omega(x_{i+1}) = \Omega^*(x_{i+1})$, the linearizing the
exponential function in terms of $d\Omega/dx$, and afterwards
analyzing the relationship to the last reversion point
\eqref{eq:3:8}, we suggest
\begin{equation}
\Omega^*(x) = 1 - \frac{\sigma}{F_c} \exp \Bigl(0.6 \ln \Bigl(
\frac{F_c}{F_c-F_i}\Bigr) \Bigr) \, x = 1 - K \bigl( \sigma, F_c,
F_i \bigr) \, x. \label{eq:3:11}
\end{equation}
The exponential term $\Omega$ and its linear approximation
$\Omega^*$ are shown in Figure \ref{fig:4}, for different
parameter ratios $\sigma/F_c^{(a,b,c)}= \{1, 2, 8\}$. Note that
the different $\Omega^*$-slopes, each one touching the same
exponential curve $\Omega^{(a,b,c)}$, correspond to different
initial states $F_i = -F_c \times \{0.2, 0.4, \ldots, 1\}$.
Since the $\Omega(x)$ term is reproduced sufficiently accurately
by $\Omega^*(x)$, and that for each following inversion state
$x_{i+1}$, we substitute \eqref{eq:3:11} into \eqref{eq:3:10}
instead of $\Omega$, making it explicitly solvable. Following to
that and solving
\begin{equation}
\frac{F_c^2}{\sigma} \Bigl( 1 - K \bigl( \sigma, F_c, F_i \bigr)
\,x \Bigr) +  F_c x -\frac{F_c^2 }{\sigma} = E_{f+}(x_i) =
E_p(F_i) \label{eq:3:12}
\end{equation}
with respect to the unknown $x$, one can calculate
\begin{equation}
x_{i+1}(F_i) = \frac{E_p(F_i)}{F_c} \, \Biggl( 1 -
\frac{F_c}{\sigma} \Bigl( \frac{F_c}{F_c - F_i} \Bigr)^{3/5}
\Biggr)^{-1}, \label{eq:3:13}
\end{equation}
where $E_p(F_i)$ is determined by \eqref{eq:3:9}. With this
intermediate result, one can directly evaluate the recovered
potential energy in the next reversion point, i.e. $E_p(i+1)$,
i.e. using \eqref{eq:3:7} with integration limits $x_0$ and
\eqref{eq:3:13}. For visualizing the development of the next
reversal, in terms of the corresponding $(x_{i+1},F_{i+1})$ state,
the force-displacement curves between two reversion points $i$ and
$i+1$ are depicted in Figure \ref{fig:5}, for the various Coulomb
friction coefficients $F_c = \{1, 1.5, 2\}$.
One can recognize that even though $|x_{i+1}| > |x_{i}|$ here,
which however depends on the $\sigma/F_c$ parametric ratio, the
$|F_{i+1}/F_{i}| < 1$ is always decreasing, thus resulting in an
always valid relationship $E_p(i+1) < E_p(i)$.

\begin{figure}[!h]
\centering
\includegraphics[width=0.9\columnwidth]{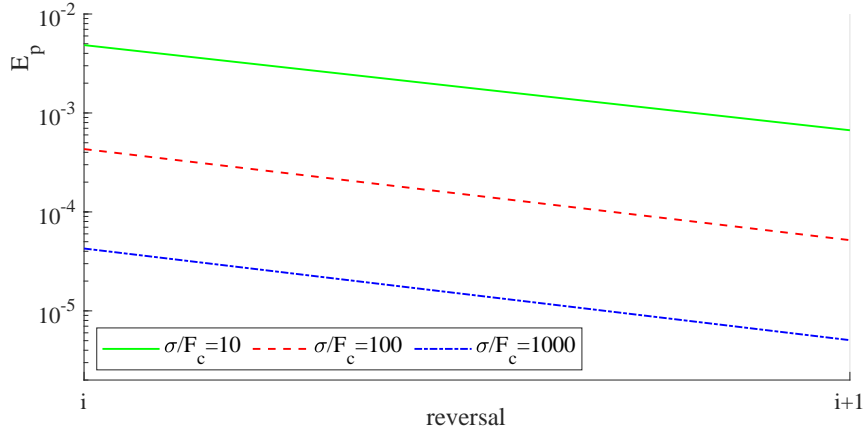}
\caption{Potential energy $E_p(i+n)$ between two consecutive
reversals, for different parametric ratios $\sigma/F_c = \{10,
100, 1000\}$.} \label{fig:6}
\end{figure}

\begin{figure}[!h]
\centering
\includegraphics[width=0.9\columnwidth]{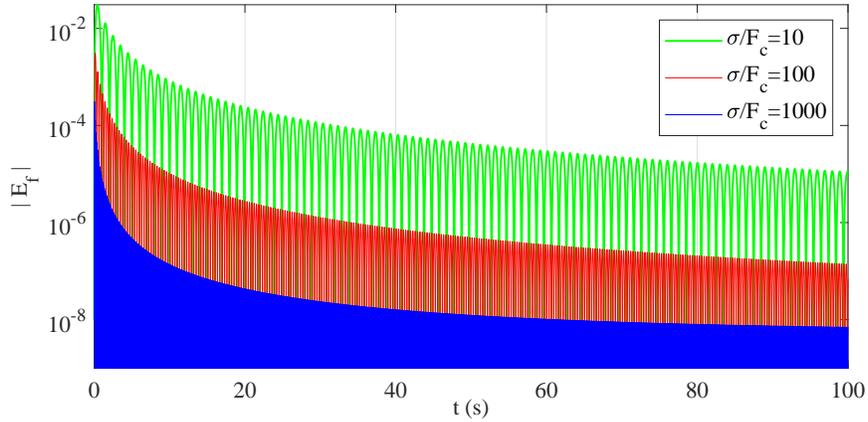}
\caption{Time series of the energy magnitude from the numerical
simulation of \eqref{eq:2:1} with \eqref{eq:3:4}, for different
parametric ratios $\sigma/F_c$.} \label{fig:7}
\end{figure}

Now, evaluating the potential energy at the following $(i+1)$-th
reversal, i.e. computing $E_p(F_{i+1})$ for the next recursive
step, cf. \eqref{eq:2:10}, we obtain
\begin{equation}
E_p(i+1) = \frac{F_c^2}{\sigma} \Biggl( \ln \Bigl( \frac{F_c}{F_c
- F_{i+1} } \Bigr) - \frac{F_{i+1}}{F_c} \Biggr). \label{eq:3:14}
\end{equation}
One can recognize that \eqref{eq:3:14} allows for an explicit
recursive form, so that it is always possible to compute
$E_p(i+n)$ for all $i < n \leq \infty$ and the given initial state
$(x_{i},F_{i})$. Recall that $E_p(i) - E_p(i+1) \equiv E_d(i)$,
which is the dissipated energy between two consecutive reversals
i.e. on one half of the hysteresis cycle. A decrease of potential
energy $E_p(i)$ (on the logarithmic scale) between two consecutive
reversals is shown in Figure \ref{fig:6}, for different parametric
ratios $\sigma/F_c = \{10, 100, 1000\}$.
One can recognize an exponential drop of $E_p(i)$ and, thus, an
exponential rate of $E_d(i+1)$. Recall that the exponential rate
of $E_d(i+1)$ and hence the convergence of the $x$ state is not in
the time series but in the series of consecutive reversals of a
decreasing hysteresis cycle. For a further growing number of
reversals $n$, the energy dissipation and the corresponding
convergence rate will result in a convergence slower than the
exponential, as can be seen from Figure \ref{fig:7}.
Here, the total energy of restoring force $E_f$ is evaluated by
the integrated power, cf. \eqref{eq:2:8}, when numerically
simulating the time series of \eqref{eq:2:1} with \eqref{eq:3:4}.
The increasing frequency of the hysteresis cycles is visible for
an increasing parametric ratio $\sigma/F_c$. One can also
recognize that the envelope of the energy peak points, which are
corresponding to the periodic reversals, has the same principal
shape for all $\sigma/F_c$ ratios. Higher $\sigma/F_c$ ratios have
an inverse biasing effect on the supposedly hyperexponential shape
of convergence and provide a faster energy dissipation during some
initial reversal cycles.

\section{Conclusions}\label{sec4}

We considered the appearance of hysteresis cycles and the
associated energy balance during the pre-sliding transients of
kinetic friction. When an unforced second-order motion system has
non-zero initial velocity, the structural hysteresis damping of
contact surface asperities gives rise to largely decreasing
nonlinear oscillations around zero equilibrium. Such behavior
occurs alongside the so-called reversal cycles, where some
restoring force energy is stored as potential energy but decreases
with each successive reversal state. Supposedly, the associated
energy dissipation and convergence of the state trajectories are
neither exponential, i.e. viscous type, nor finite-time, i.e.
Coulomb type.

We have analyzed such pre-sliding hysteresis cycles, being
inspired by the pervious related works
\cite{bliman1992mathematical}, \cite{AlBender2004}. We established
an energy balance between the potential energy associated with
each pair of two consecutive reversion states and the dissipated
hysteretic energy during the half-cycle. In particular, we
demonstrated a recursive series expansion of dissipated energies
for an infinite sequence of reversion points and showed that this
series converges to the upper bounded potential energy in the
first (initial) reversal point. The analysis developed led to
expressions for energies and force-displacement curves that allow
application to various hysteresis models. The class of the
suitable force-displacement mapping is limited to the
rate-independent clockwise hysteresis functions.

As an illustrative example, we considered the Dahl model
\cite{dahl1968solid} of the solid Coulomb friction without
discontinuities at the velocity zero-crossing. We developed an
explicit approximation for the energies and states of the
consecutive reversals, that allows analyzing the energy
dissipation and state trajectories depending on the system
parameters -- the Coulomb friction coefficient and the so-called
rest stiffness. This is also helpful for further convergence
analysis and investigation of the motion stop in presence of
Coulomb friction without discontinuity. Our analysis is in line
with the original results demonstrated in \cite{dahl1976}, while
providing an accurate estimation of the restoring force energy per
half-cycle after each force zero-crossing, cf. with \cite[Fig.
7]{dahl1976}. The discussed energy dissipation by the Dahl model
discloses further its main features and properties. Only
asymptotic convergence of the motion state trajectories is
possible with use of the Dahl model, cf. Figure \ref{fig:7}. Thus,
the corresponding kinetic energy tends towards zero with time
towards infinity. This reveals some of the model's weaknesses in
relation to a more realistic and physically reasoned behavior of
the pre-sliding kinetic friction.

The results of this study are applicable to other types of kinetic
friction modeling with hysteresis and should contribute to a
better understanding and prediction of the pre-sliding
force-displacement properties, and the associated energy
dissipation on the rubbing and slipping contact interfaces.

{\small

\bibliographystyle{amsplain}        
\bibliography{references}

%

}

{\small {\em Authors' addresses}: {\em Michael Ruderman},
University of Agder, Kristiansand, Norway
\\ e-mail: \texttt{michael.ruderman@uia.no}.

}

\end{document}